documentstyle[preprint,aps]{revtex}
\documentstyle[preprint,aps]{revtex}
\begin{document}
\draft

\title{Pseudoclassical description of the massive Dirac 
particles in odd dimensions}
\author{D.M. Gitman}

\address{Instituto de F\'{\i}sica, Universidade de S\~ao Paulo\\ 
P.O. Box 66318, 05389-970 S\~ao Paulo, SP, Brazil}

\author{A.E. Gon\c calves}
\address{Departamento de F\'{\i}sica, Universidade Estadual de Londrina\\ 
P.O. Box 6001, 86051-970 Londrina, Pr, Brazil}
\date{\today}
\maketitle
\begin{abstract}
A  pseudoclassical model is proposed to describe massive Dirac  
( spin one-half) particles in arbitrary odd dimensions. The quantization of 
the model reproduces  the minimal quantum theory of spinning particles 
in such dimensions. A dimensional duality between the model proposed and the 
pseudoclassical description of Weyl particles in even dimensions is discussed.
\end{abstract}

\section{Introduction}
As it is known  one can construct a pseudoclassical model 
to describe massive Dirac (spin one-half) particles in $3+1$ dimensions \cite{BM}. 
Its generalization to the case of even  dimensions  $D=2n\;,\;\;n=3,\;4,\ldots ,$ 
can be done by means of the direct dimensional extension
\cite{GG}. The corresponding action has the form
 
\begin{equation}
\displaystyle
S=\int_0^1\left[-\frac{\dot{x}^2}{2e}-e\frac{m^2}{2}+\imath
\left(\frac{\dot{x}_\mu\psi^\mu}{e}-m\psi^D\right)\chi -
\imath\psi_a\dot{\psi}^a\right]d\tau\;,
\end{equation}
where $\dot{x}^2=\dot{x}_\mu\dot{x}^\mu$; the Greek (Lorentz) indices 
$\mu$, $\nu$, $\ldots$, run over 0, 1, $\ldots\;,\;d-1$, whereas the
Latin ones $a,\;b$, run over 0, 1, $\ldots\;,\;d$;  
$\eta_{\mu\nu}=\hbox{diag}(\underbrace{1,-1,\ldots, -1}_D)$, 
$\eta_{ab}=\hbox{diag}(\underbrace{1,-1,\ldots ,-1}_{D+1})$. The variables 
$x^\mu$ and $e$ are even and $\psi^n$, $\chi$ are odd. The quantization 
of the model leads to the Dirac quantum theory of spin one-half particle 
(to the Dirac equation).

Attempts to extend the pseudoclassical description 
to the arbitrary odd-dimensional case had met some problems, which are 
connected with the absence of an analog of $\gamma^5-$matrix in such 
dimensions. For instance, in $2n+1$ dimensions the direct generalization 
of the standard action \cite{BM} does not reproduce a minimal quantum 
theory of spinning particle, where particles with spin $1/2$ and $-1/2$ 
have to be considered as different ones. In the papers \cite{P} 
two modifications of the standard action were proposed to solve the 
problem. From our point of view both have essential shortcomings to 
believe the problem is closed. For instance, the first action \cite{P} 
is classically equivalent to the standard action and does not provide 
required quantum properties in course of canonical and path-integral 
quantization. Moreover, it is $P-$ and $T-$ invariant, so that an anomaly 
is present. Another one \cite{P} does not obey gauge supersymmetries and 
therefore loses the main attractive feature in such kind of models, which 
allows one to treat them as prototypes of superstrings or some modes in 
superstring theory. In \cite{GGT0} a new pseudoclassical model for a 
massive Dirac particle in $2+1$ dimensions was proposed, which obeys all 
the necessary symmetries, is $P-$ and $-T$ non-invariant and reproduces the 
minimal quantum theory of the Dirac particle in $2+1$ dimensions. It 
turns out to be possible to generalize this model to arbitrary 
odd-dimensional case. We present such a generalization in the present 
paper. First, we consider the hamiltonization of the theory and its 
quantization. Then we discuss a remarkable dimensional duality between the 
model proposed and the pseudoclassical description of massless spinning 
particles in even dimensions.
\section{Pseudoclassical description}
 In odd dimension $D=2n+1$ we propose the 
following action to describe spinning particles 
\begin{eqnarray}\label{act}
\displaystyle
S &=&\int\limits_0^1\left[-\frac{z^2}{2e}-e\frac{m^2}{2}-
\imath m\psi^{2n+1}\chi-\frac{s}{2^n}m\kappa-
\imath\psi_a\dot\psi^a\right]d\tau\nonumber\\ 
\displaystyle
&\equiv &\int\limits_0^1Ld\tau\;,\;\;
z^\mu=\dot x^\mu-\imath\psi^\mu\chi+\frac{(2\imath)^n}{(2n)!}
\varepsilon^{\mu\rho_1\ldots\rho_{2n}}\psi_{\rho_1}\ldots 
\psi_{\rho_{2n}}\kappa\;;
\end{eqnarray}
Here a new even variable $\kappa$ is introduced and 
$\varepsilon^{\mu\nu\ldots\lambda}$ is Levi-Civita tensor density in 
$2n+1$-dimensions normalized by $\varepsilon^{01,\ldots ,2n}=1,\;\;s$
is an even constant of the Berezin algebra. 
We suppose that $x^\mu$ and $\psi^\mu$ are Lorentz vectors and $e$,
$\kappa$, $\psi^{2n+1}$, $\chi$ are scalars so that the action (\ref{act}) 
is invariant under the restricted Lorentz transformations 
(but not $P$- and $T$-invariant). There are three types of gauge 
transformations, under which the action (\ref{act}) is invariant:  
reparametrizations
\begin{equation}\label{rep}
\delta x^\mu=\dot x^\mu\xi\;,\;\; \delta e=\frac{d}{d\tau}(e\xi)\;,
\;\;\delta\psi^a=\dot\psi^a\xi\;\;,
\delta\chi=\frac{d}{d\tau}(\chi\xi)\;, \;\;
\delta\kappa=\frac{d}{d\tau}(\kappa\xi)\;, 
\end{equation}
with an even parameter $\xi(\tau)$; supertransformations
\begin{equation}\label{st1}
\delta x^\mu=\imath\psi^\mu \epsilon\;,\;\;\delta e=\imath\chi\epsilon\;\;,
\delta\psi^\mu=\frac{z^\mu}{2e}\epsilon\;\;,
\delta\psi^{2n+1}=\frac{m}{2}\epsilon,\;\;\delta\chi=\dot\epsilon\;,
\;\;\delta\kappa=0\;, 
\end{equation}
with an odd parameter $\epsilon(\tau)$; and additional supertransformations
\begin{eqnarray}\label{st2}
&&\delta x^\mu=-\frac{(2\imath)^n}{(2n)!}\varepsilon^{\mu\rho_1\ldots\rho_{2n}}
\psi_{\rho_1}\ldots\psi_{\rho_{2n}}\theta\;,\nonumber\\
&&\delta\psi^\mu=
-\frac{\imath}{e}\frac{(2\imath)^n}{(2n)!}
\varepsilon^{\mu\rho_1\ldots\rho_{2n}}z_{\rho_1}\psi_{\rho_2}
\ldots\psi_{\rho_{2n}}\theta\;,\nonumber\\
&&\delta\kappa=\dot\theta + \frac{s}{m}
\frac{2^{2n+1}\imath^n(n-1)z_\mu}{(2n)!e}
\varepsilon^{\mu\rho_1\ldots\rho_{2n}}\dot\psi_{\rho_1}\psi_{\rho_2}\ldots
\psi_{\rho_{2n}}\theta\;,\nonumber\\
&&\delta e=\delta\psi^{2n+1}=\delta\chi=0\;,
\end{eqnarray}
with an even parameter $\theta(\tau)$. 

The total angular momentum tensor $M_{\mu\nu}$, is
\begin{equation}\label{am}
M_{\mu\nu}=x_\mu\pi_\nu -x_\nu\pi_\mu +\imath [\psi_\mu ,\psi_\nu ]\;,
\end{equation}
where $\pi_\nu = \partial L/\partial{\dot x}^\nu$. 

Going over to the Hamiltonian formulation, we introduce the canonical momenta
\begin{eqnarray}\label{cm}
\displaystyle
&&\pi_\mu=\frac{\partial L}{\partial\dot x^\mu}=-\frac{1}{e}z_\mu\;,\;\;
P_e=\frac{\partial L}{\partial\dot e}=0\;, \;\;
P_\chi=\frac{\partial_rL}{\partial\dot\chi}=0\;, \nonumber\\
\displaystyle
&&P_\kappa=\frac{\partial L}{\partial\dot\kappa}=0\;,\;\;
P_a=\frac{\partial_rL}{\partial\dot\psi^a}=-\imath\psi_a\;.
\end{eqnarray}
It follows from (\ref{cm}) that there exist primary constraints
\begin{equation}\label{pc}
\Phi^{(1)}_1=P_e\;, \;\; \Phi^{(1)}_2=P_\chi\;, \;\;
\Phi^{(1)}_3=P_\kappa\;,\;\; \Phi^{(1)}_{4a}=P_a+\imath\psi_a\;.
\end{equation}
Constructing the total Hamiltonian $H^{(1)}$, according to the standard
procedure \cite{D,GT1}, we get $H^{(1)}=H+\lambda_A\Phi^{(1)}_A$, where 
\begin{eqnarray}\label{ham}
H=&-&\frac{e}{2}(\pi^2-m^2)+
\imath(\pi_\mu\psi^\mu+m\psi^{2n+1})\chi \nonumber\\ 
&-&\left[\frac{(2\imath)^n}{(2n)!}
\varepsilon^{\mu\rho_1\ldots\rho_{2n}}\pi_\mu\psi_{\rho_1}
\ldots\psi_{\rho_{2n}}
-\frac{1}{2^n}sm \right]\kappa\;.
\end{eqnarray}
From the consistency conditions $\dot\Phi^{(1)}=\{\Phi^{(1)},H^{(1)}\}=0$ we
find secondary constraints $\Phi^{(2)}=0$,
\begin{eqnarray}\label{sc}
\displaystyle
&&\Phi^{(2)}_1=\pi_\mu\psi^\mu+m\psi^{2n+1}\;,\;\; \Phi^{(2)}_2 =
\pi^2-m^2\;,\nonumber\\
&&\Phi^{(2)}_3= 
\frac{(2\imath)^n}{(2n)!}
\varepsilon^{\mu\rho_1\ldots\rho_{2n}}\pi_\mu\psi_{\rho_1}
\ldots\psi_{\rho_{2n}}
-\frac{1}{2^n}sm\;,
\end{eqnarray}
and determine $\lambda$, which correspond to the primary constraints
$\Phi^{(1)}_{4}$. No more secondary constraints arise from the
consistency conditions and the Lagrangian multipliers, correspondent to the
primary constraints $\Phi^{(1)}_i$, $i=1,2,3$, remain undetermined. The
Hamiltonian (\ref{ham}) is proportional to the constraints as one
could expect in the case of a reparametrization invariant theory. 
One can go over from the initial set of constraints $\Phi^{(1)},
\Phi^{(2)}$ to the equivalent ones $\Phi^{(1)},\tilde{\Phi}^{(2)}$,
where $\tilde{\Phi}{}^{(2)}=\Phi^{(2)}\left(\psi\rightarrow\tilde{\psi}=\psi+
\frac{\imath}{2}\Phi^{(1)}_4 \right)$.  
The new set of constraints can be explicitly divided in a set of the
first-class constraints, which are ($\Phi^{(1)}_i$, $i=1,2,3$,
$\tilde{\Phi}^{(2)}$) and in a set of second-class constraints
$\Phi^{(1)}_4$. Thus, we are dealing with a theory with first-class
constraints.
\section{Quantization}
Let us consider first the Dirac quantization, where the second-class
constraints define the Dirac brackets and therefore the commutation relations,
whereas, the first-class constraints, being applied to the state vectors,
define physical states. For essential operators and nonzeroth
commutation relations one can obtain in the case of consideration: 
\begin{equation}\label{cr}
[\hat{x}^\mu,\hat{\pi}_\nu]=\imath\{x^\mu,\pi_\nu\}_{D(\Phi^{(1)}_4)}=
\imath\delta^\mu_\nu\;, \;\;
[\hat{\psi}^a,\hat{\psi}^b]_+=\imath\{\psi^a,\psi^b\}_{D(\Phi^{(1)}_4)}=
-\frac{1}{2}\eta^{ab}\;. 
\end{equation}
It is possible to construct a realization of the commutation relations
(\ref{cr}) in a Hilbert space ${\cal R}$ whose elements ${\bf f}\in 
{\cal R}$ are 
$2^{n+1}$ component columns dependent on $x$,
\begin{equation}\label{fcc}
{\bf f}(x)=\left(\begin{array}{c}u_-(x)\\u_+(x)\end{array}\right),
\end{equation}
where $u_\mp(x)$ are $2^n$ component columns. Then
\begin{equation}\label{xpr}
\hat{x}^\mu=x^\mu{\bf I}\;,\;\; \hat{\pi}_\mu=-\imath\partial_\mu{\bf I}\;,
\;\; \hat{\psi}^a=\frac{\imath}{2}\gamma^a, 
\end{equation}
here ${\bf I}$ is
$2^{n+1}\times 2^{n+1}$ unit matrix and $\gamma^a$, 
$a=0,1,\ldots,2n+1$ are $\gamma$-matrices
in 2(n+1)-dimensions \cite{Case}, which we select in the spinor representation
$\gamma^0={\rm antidiag}(I,\;I),\;\;\gamma^i={\rm antidiag}
(\sigma^i,\;-\sigma^i),\;\; i=1,2,\ldots , 2n+1\;,$ 
where $I$ is $2^n\times 2^n$ unit matrix, and $\sigma^i$ are $2^n\times 2^n$ 
$\sigma$-matrix, which obey the Clifford algebra, $[\sigma^i,\sigma^j]_+=2
\delta^{ij}$.

According to the scheme of quantization selected, the operators of the 
first-class constraints have to be applied to the state vectors to
define physical sector, namely, $\hat{\Phi}^{(2)}{\bf f}(x)=0\;$, 
where $\hat{\Phi}^{(2)}$ are operators, which correspond to the constraints
(\ref{sc}).  There is no ambiguity in the construction of the operator
$\hat{\Phi}^{(2)}_1$ according to the classical function $\Phi^{(2)}_1$.
Taken into account the realization (\ref{fcc}), (\ref{xpr}), one can
present the equations $\hat{\Phi}^{(2)}{\bf f}(x)=0\;$  in the 
$2^n$-component form, 
\begin{equation}\label{setc}
\displaystyle
[\imath\partial_\mu\gamma^\mu-m\gamma^{2n+1}]{\bf f}(x)=0 \;\;
\Longleftrightarrow\left\{ 
\begin{array}{c}
[\imath\partial_\mu\Gamma^\mu_+-m]u_+(x)=0\;, \nonumber\\
\displaystyle
[\imath\partial_\mu\Gamma^\mu_-+m]u_-(x)=0\;,
\end{array}\right.
\end{equation}
where two sets of $\gamma$-matrices $\Gamma^\mu_{\varsigma},\;
\varsigma =\pm$, in $2n+1$ 
dimensions are introduced,
\begin{eqnarray}
\displaystyle
&&\Gamma^0_\varsigma=\sigma^{2n+1}, \;\; \Gamma^1_\varsigma=
\varsigma\sigma^{2n+1}
\sigma^1,\ldots\;,
\Gamma^{2n}_\varsigma=\varsigma\sigma^{2n+1}\sigma^{2n}\;,
\nonumber\\
\displaystyle
&&\Gamma^\mu_-=\Gamma_{+\mu}\;, \;\;
[\Gamma^\mu_\varsigma ,\Gamma^\nu_\varsigma]_+=2\eta^{\mu\nu}\;.
\end{eqnarray}

The is a relation $\hat{\Phi}^{(2)}_2=(\hat{\Phi}^{(2)}_1)^2$ so that 
the equation $\hat{\Phi}^{(2)}_2{\bf f}=0$ is not independent. 
The equation $\hat{\Phi}^{(2)}_3{\bf f}(x)=0$ can be 
presented in the following form 
\[
\left[\frac{(-\imath)^n}{(2n)!}
\varepsilon^{\mu\rho_1\ldots\rho_{2n}}(\imath\partial_\mu)
\gamma_{\rho_1}\ldots\gamma_{\rho_{2n}}+sm\right]{\bf f}(x)=0 \;\;
\]
or in $2^n$-component form 
\begin{eqnarray}\label{ophi3}
\displaystyle
&&[\imath\partial_\mu\Gamma^\mu_++(-1)^nsm]u_+(x)=0\;, \nonumber\\
\displaystyle
&&[\imath\partial_\mu\Gamma^\mu_-+(-1)^nsm]u_-(x)=0\;.
\end{eqnarray}
In quantum theory one has to select $s=\pm 1$, then, combining eq. 
(\ref{setc}) and (\ref{ophi3}), we get 
\begin{equation}
[\imath\partial_\mu\Gamma^\mu_s-\varsigma m]u_{\varsigma}(x)=0\;,\;\;
u_{-\varsigma}(x)\equiv 0\;,\;\;\varsigma=(-1)^ns=\pm 1\;.
\end{equation}
To interpret the result obtained one has to calculate also the operators
$\hat{M}_{\mu\nu}$ correspondent to the angular momentum tensor (\ref{am}),
\[
\hat{M}_{\mu\nu}=-\imath(x_\mu\partial_\nu-x_\nu\partial_\mu)-\frac{\imath}{4}
\left(\begin{array}{cc}[\Gamma_{-\mu},\Gamma_{-\nu}]&0\\
0&[\Gamma_{+\mu},\Gamma_{+\nu}]\end{array}\right)\;.
\]
Thus, in the quantum mechanics constructed, the states with $\varsigma =+$ are
described by the $2^n$-component wave function $u_+(x)$, which obeys the
Dirac equation in $2n+1$ dimensions and is transformed under the Lorentz 
transformation as spin $+1/2$.
For $\varsigma =-$ the quantization leads to the theory of $2n+1$ Dirac 
particle with spin $-1/2$ and the wave function $u_-(x)$.

To quantize the theory canonically we have to impose as much as possible
supplementary gauge conditions to the first-class constraints. 
In the case under consideration, it turns out to be possible to 
impose gauge conditions to all the first-class constraints, excluding 
the constraint $\tilde{\Phi}^{(2)}_3$. Thus, we are fixing the gauge 
freedom, which  corresponds to two types of gauge transformations 
(\ref{rep}) and (\ref{st1}).  As a result  we remain only with one 
first-class constraint, which is a reduction of
$\Phi^{(2)}_3$ to the rest of constraints and gauge conditions. It can be
used to specify the physical states. All the second-class constraints 
form the Dirac brackets. 
The following gauge conditions $\Phi^G=0$ are imposed: 
$\Phi^G_1=e+\zeta\pi^{-1}_0\;,\;\; \Phi^G_2=\chi\;,\;\; \Phi^G_3=\kappa\;,\;\;
\Phi^G_4=x_0-\zeta\tau\;,\;\; \Phi^G_5=\psi^0\;,$ 
where $\zeta=-\hbox{sign}\,\pi^0$. (The gauge $x_0-\zeta\tau=0$ was
first proposed in \cite{GT1,GT2} as a conjugated gauge condition to
the constraint $\pi^2-m^2=0$). Using the consistency condition
$\dot\Phi^G=0$, one can determine the Lagrangian multipliers,
which correspond to the primary constraints $\Phi^{(1)}_i$,
$i=1,2,3$. 
To go over to a time-independent set of constraints (to use standard
scheme of quantization without any modifications \cite{GT1}) 
we introduce the variable $x^\prime_0,\;x^\prime_0=x_0-\zeta\tau$, 
instead of $x_0$, without changing the rest of the variables. 
That is a canonical transformation in the space of all variables with 
the generating function $W=x_0\pi^\prime_0+\tau|\pi^\prime_0|+W_0$, 
where $W_0$ is the generating function of the identity transformation 
with respect to all variables except $x^0$ and $\pi_0$. 
The transformed Hamiltonian $H^{(1)\prime}$ is of the form
\begin{equation}\label{ph1}
H^{(1)\prime}=H^{(1)}+{\partial W\over\partial\tau}=\omega+\{\Phi\}\;,\;\;
\omega=\sqrt{\pi^2_d+m^2}\;, \;\;\; d=1,2,\ldots ,2n\;, 
\end {equation}
where $\{\Phi\}$ are terms proportional to the constraints and
$\omega$ is the physical Hamiltonian.
All the constraints of the theory, can be presented after this
canonical transformation in the following equivalent form:  
$K=0$, $\phi=0$, $T=0$, where
\begin{eqnarray}\label{cc}
\displaystyle
&&K=(e-\omega^{-1}\;,\; P_e\;;\;\; \chi\;,\; P_\chi\;; \;\; \kappa\;,\; 
P_\kappa\;;\;\;
x^\prime_0\;,\; |\pi_0|-\omega\;; \;\; \psi^0\;,\; P_0)\;; \nonumber\\ 
\displaystyle
&&\phi =(\pi_d\psi^d+m\psi^{2n+1}, P_k+\imath\psi_k)\;,\;\;
k=1,2,\ldots ,2n+1\;;\nonumber\\
&&T=\frac{(2\imath)^n}{(2n)!}
\zeta\omega\varepsilon^{i_1\ldots i_{2n}}\psi_{i_1}\ldots 
\psi_{i_{2n}}+\frac{sm}{2^n}\;,\;\;i_d=1,2,\ldots ,2n\;.
\end{eqnarray}
The constraints $K$ and $\phi$ are of the second-class, whereas $T$ is the
first-class constraint.  Besides, the set $K$ has the so called special form
\cite{GT1}. In this case, if we eliminate the variables $e$, $P_e$, $\chi$,
$P_\chi$, $\kappa$, $P_\kappa$, $x^\prime_0$, $|\pi_0|$, $\psi^0$, and $P_0$,
using the constraints $K=0$, the Dirac brackets with respect to all
the second-class constraints $(K,\phi)$ reduce to ones
with respect to the constraints $\phi$ only. Thus, on this stage, we will
only consider the variables $x^d$, $\pi_d$, $\zeta$, $\psi^k$, $P_k$ and two
sets of constraints - the second-class ones $\phi$ and the first-class one
$T$. Nonzeroth Dirac brackets for the independent variables are
\begin{eqnarray}\label{db}
\displaystyle
&&\{x^d,\pi_r\}_{D(\phi)}=\delta^d_r\;, \;\;
\{x^d,x^r\}_{D(\phi)}=\frac{\imath}{\omega^2}[\psi^d,\psi^r]\;, \;\;
\{x^d,\psi^r\}_{D(\phi)}=-\frac{1}{\omega^2}\psi^d\pi_r\;, \nonumber\\ 
\displaystyle
&&\{\psi^d,\psi^r\}_{D(\phi)}=-\frac{\imath}{2}
(\delta^d_r-\omega^{-2}\pi_d\pi_r)\;,
\;\;d,r=1,2,\ldots ,2n\;.
\end{eqnarray}
Going over to the quantum theory, we get the commutation
relations between the operators $\hat{x}^d$, $\hat{\pi}_d$, $\hat{\psi}^d$ by
means of the Dirac brackets (\ref{db}),
\begin{eqnarray}\label{rc}
\displaystyle
&&[\hat{x}^d,\hat{\pi}_r]=\imath\delta^d_r\;,\;\;
[\hat{x}^d,\hat{x}^r]=-\frac{1}{\hat{\omega}^2}[\hat{\psi}^d,\hat{\psi}^r]\;,
\nonumber\\
\displaystyle
&&[\hat{x}^d,\hat{\psi}^r]=-\frac{\imath}{\hat{\omega}^2}\hat{\psi}^d
\hat{\pi}_r\;,\;\;[\hat{\psi}^d,\hat{\psi}^r]_+=\frac{1}{2}(\delta^d_r-
\hat{\omega}^{-2}\hat{\pi}_d\hat{\pi}_r)\;.
\end{eqnarray}
We assume as usual \cite{GT1,GT2} the operator $\hat{\zeta}$ to have 
the eigenvalues
$\zeta=\pm1$ by analogy with the classical theory, so that $\hat{\zeta}^2=1$,
and also we assume the equations of the second-class constraints 
$\hat{\phi}=0$. Then one can
realize the algebra (\ref{rc}) in a Hilbert space ${\cal R}$, 
whose elements ${\bf f}\in {\cal R}$
are $2^{n+1}$ component columns dependent on ${\bf x}=(x^d)$, 
$d=1,2,\ldots ,2n$,
\begin{equation}\label{fcc1}
{\bf f}({\bf x})=\left(\begin{array}{c}f_+({\bf x})\\f_-({\bf x})
\end{array}\right)\;, 
\end{equation}
so that $f_+({\bf x})$ and $f_-({\bf x)}$ are $2^n$ component columns. A
realization of the commutations relations has the form
\begin{eqnarray}\label{r}
\displaystyle
&&\hat{x}^d=x^d{\bf I}-\frac{\imath}{4\hat{\omega}^2}[\Sigma^d,
\hat{\pi}_r\Sigma^r]_- -
\frac{\imath m}{4\hat{\omega}^2}[
\Sigma^d, \Sigma^{2n+1}]_-\;,\;\;\hat{\pi}_r=-\imath\partial_r{\bf I}\;,
\nonumber \\ 
\displaystyle
&&\hat{\psi}^d=\frac{1}{2}\left(
\delta^d_r-\hat{\omega}^{-2}\hat{\pi}_d\hat{\pi}_r\right)
\Sigma^r-\frac{m\hat{\pi}_d }{2\hat{\omega}^2}\Sigma^{2n+1}, \;\;
\hat{\zeta}=\left(\begin{array}{cc}I&0\\0&-I\end{array}\right),
\end{eqnarray}
where ${\bf I}$ and $I$ are $2^{n+1}\times 2^{n+1}$  and 
$2^n\times 2^n$ unit matrices,
$\Sigma^k=\hbox{diag}( \sigma^k,\sigma^k)$. 
The operator $\hat{T}$ correspondent to the first-class constraint 
$T$ (see (\ref{cc})) appears to be 
\begin{equation}
\hat{T}=\frac{\varsigma m}{\hat{\omega}}\hat{\zeta}\Sigma^{2n+1}\left[
\hat{\zeta}\hat{\omega}\Sigma^{2n+1}+\imath\partial_d
\left(\varsigma\Sigma^{2n+1}\Sigma^d\right)-\varsigma m\right]\;,\;\;
\varsigma = (-1)^ns=\pm 1\;.
\end{equation}
The latter operator specifies the physical states according to scheme
of quantization selected, $\hat{T}{\bf f}=0$. 
On the other hand, the state vectors ${\bf f}$ have to obey the Schr\"odinger
equation, which defines their ``time'' dependence, 
$(\imath\partial/\partial\tau-\hat{\omega}){\bf f}=0\;,\;\;
\hat{\omega}=\sqrt{\hat{\pi}^2_d+m^2}$, where the quantum
Hamiltonian $\hat{\omega}$ corresponds the classical one $\omega$ (\ref{ph1}). 
Introducing the
physical time $x^0=\zeta\tau$ instead of the parameter $\tau$ \cite{GT2,GT1},
we can rewrite the Schr\"odinger equation in the following form 
(we can now write ${\bf f}={\bf f}(x),\;(x=x^0,\;{\bf x})$), 
\begin{equation}\label{se}
(\imath\frac{\partial}{\partial x^0}-\hat{\zeta}\hat{\omega}){\bf f}(x)=0\;. 
\end{equation}
Using (\ref{se}) in the eq. $\hat{T}{\bf f}=0$, namely replacing there
the combination 
$\hat{\zeta}\hat{\omega}{\bf f}$ by $\imath\partial_0{\bf f}$, one can verify
that both components $f_\pm (x)$, of the state vector (\ref{fcc1}) obey
one and the same equation
\begin{equation}
(\imath\partial_\mu\Gamma^\mu_{\varsigma}-\varsigma m)f_\zeta(x)=0\;,\;\;
\zeta =\pm 1\;,
\end{equation}
which is the $2n+1$ Dirac equation for a particle of spin $\varsigma/2$ 
whereas $f_\pm(x)$ can be interpreted (taken into account (\ref{se})) as
positive and negative frequency solutions to the equation
respectively. Substituting the realization (\ref{r}) into the
expression (\ref{am}), we get the generators of the Lorentz transformations
\begin{equation}
\hat{M}_{\mu\nu}=-\imath(x_\mu\partial_\nu-x_\nu\partial_\mu)-\frac{\imath}{4}
\left(\begin{array}{cc}[\Gamma_{\varsigma\mu},\Gamma_{\varsigma\nu}]&0\\
0&[\Gamma_{\varsigma\mu},\Gamma_{\varsigma\nu}]\end{array}\right),
\end{equation}
which have the standard form for both components $f_\zeta (x)$. Thus, a natural
interpretation of the components $f_\zeta(x)$ is the following: $f_+(x)$ is
the wave function of a particle with spin $\varsigma/2$ and $f^*_-(x)$ is 
the wave function of an anti-particle with spin $\varsigma/2$. 

\section{Dimensional duality between massive and massless spinning particles}
\label{Ddb}
As is known, the method of dimensional reduction \cite{DNP} appears to 
be often useful to construct models (actions) in low dimensions using 
some appropriate models in higher dimensions. In fact, such kind of ideas 
began from the works \cite{KK}. One can also mention that the method 
of dimensional reduction was used to interpret masses in supersymmetric 
theories as components of momenta in space of higher dimensions, which 
are frozen in course of the reduction. It is interesting that the model 
of Dirac particles in odd dimensions proposed in the present paper  
is related to the model \cite{GG2} 
of Weyl particles in even dimensions by means of a dimensional reduction. 

The action and the Hamiltonian of the latter model in $D=2(n+1)$
dimensions  have the form 
\begin{eqnarray}\label{SHC}
\displaystyle
S &=&\int\limits_0^1\left[-\frac{z^2}{2e}-
\imath\psi_\mu\dot\psi^\mu\right]d\tau\;,\nonumber\\
z^\mu&=&\dot x^\mu-\imath\psi^\mu\chi+\frac{(2\imath)^{\frac{D-2}{2}}}{(D-2)!}
\varepsilon^{\mu\nu\rho_2\ldots\rho_{D-1}}b_\nu\psi_{\rho_2}\ldots 
\psi_{\rho_{D-1}}+\frac{s}{2^{\frac{D-2}{2}}}b^\mu\;.\nonumber\\
H &=& -\frac{e}{2}\pi^2+\imath\pi_\mu\psi^\mu\chi\nonumber\\ 
&&\;-\left[
\frac{(2\imath)^\frac{D-2}{2}}{(D-2)!}
\epsilon_{\nu\mu\rho_2\ldots\rho_{D-1}}\pi^\mu
\psi^{\rho_2}\ldots\psi^{\rho_{D-1}}+
\frac{\alpha}{2^\frac{D-2}{2}}\pi_\nu\right] b^\nu\;.\nonumber\\
\end{eqnarray}
In the canonical gauge similar to one which was considered above, in particular, 
in the gauge $\psi^0=0$, we remain only with the first-class constraints 
\begin{equation}\label{fcct}
T_\mu =
\frac{(2\imath)^\frac{D-2}{2}}{(D-2)!}
\epsilon_{\nu\mu\rho_2\ldots\rho_{D-1}}\pi^\mu
\psi^{\rho_2}\ldots\psi^{\rho_{D-1}}+
\frac{\alpha}{2^\frac{D-2}{2}}\pi_\nu=0\;.
\end{equation}
One can see that, in fact, among the constraints (\ref{fcct}) only one is 
independent 
\[
T_\mu =\frac{\pi_\mu}{\pi_0}T_0\;.
\]
Thus one can use only one component of $b^\mu$ and all others put to be 
zero. Now one can do a dimensional reduction $2(n+1)\rightarrow 2n+1$ 
in the Hamiltonian and constraint $T_0$, putting also $\pi_{2n+1}=m$, 
$b^{2n+1}=-\kappa$, $b^0=b^1=\ldots =b^{2n}=0$. As a result of such 
a procedure we just obtain the expressions (\ref{ham}) and (\ref{cc}) 
for the Hamiltonian and the constraint.  
The second class constraints of the model (\ref{SHC}) also coincide 
with ones of the model (\ref{act}) after the dimensional reduction. 
Thus, there exist a dimensional duality between the massive spinning 
particles in odd dimensions and massless ones in even dimensions. 

\section{Discussion}
One ought to say that at present there exist pseudoclassical models (PM) 
to describe all massive higher spins ( integer and half-integer) in 3+1 
dimensions \cite{BM,PP}. Generalization of the models to arbitrary even 
dimensions can be easily done by means of a trivial dimensional extension 
similarly to the spin one-half case. To get the PM for higher 
spins in arbitrary odd dimensions one can start
from the model proposed in the present paper, using the ideas of the work 
\cite{GT}. Namely, one has to multiplicate the variables $\psi,\; 
\chi, \; \kappa,\;s$ in the action (\ref{act}).
Then an appropriate action has the form
\begin{eqnarray}\label{S1}
S&=&\int_{0}^1 \left\{-\frac{z^2}{2e}-e\frac{m^2}{2}-\sum_{A=1}^N\left[
sm\left(\frac{\kappa_{A}}{2^n}+i\psi_{A}^{2n+1}\chi_{A}\right)+
i\psi_{Aa}\dot{\psi}_{A}^{a}\right]\right\}d\tau\;,\nonumber\\
z^\mu &=&\dot{x}^\mu - \sum_{A=1}^N\left[\imath\psi^\mu_A\chi_A -
\frac{(2\imath)^n}{(2n)!}\varepsilon^{\mu\rho_1\ldots\rho_{2n}}
\psi_{A\rho_1}\ldots\psi_{A\rho_{2n}}\kappa_A\right]\;.
\end{eqnarray}
Certainly, a detailed analysis of the action (\ref{S1}) and its quantization may
demand essential technical work in higher dimensions. In spite of in 2+1 
dimensions the model can be quantized explicitly for all higher spins
both canonically and by means of the Dirac method \cite{GT}, in 3+1 dimensions 
the corresponding PM \cite{PP} was quantized canonically only for spins 
one-half \cite{BM} and one \cite{GGT}. As to the massless particles 
spin one-half, the corresponding PM exist at present in arbitrary even 
dimensions \cite{GG2,GGT1}.
Its generalization to describe higher spins can be done in the same manner
\begin{eqnarray}\label{S2}
S&=&\int_0^1\left[-\frac{1}{2e}z^2
-\imath\sum_{A=1}^N\psi_{A\mu}\dot{\psi}^\mu_A\right]d\tau\;,
\nonumber\\
z^\mu &=&\dot{x}^\mu-\sum_{A=1}^N
\left(\imath\psi_A^\mu\chi_A-\imath\varepsilon^{\mu\nu\rho\varsigma}
b_{A\nu}\psi_{A\rho}\psi_{A\varsigma}-\frac{1}{2}sb^\mu_A\right)\;.
\end{eqnarray}

There exist the dimensional duality mentioned above between the models (\ref{S1}) and 
(\ref{S2}).

Massless higher spins in arbitrary odd dimensions can be described
pseudoclassically by the model, which follows from (\ref{S1}) in the limit 
$m\rightarrow 0$ .

Thus, at present, in principle, we have PM to describe all integer and half-
integer spins in arbitrary dimensions. 

\begin{center}
Acknowledgments
\end{center}
D.M. Gitman  thanks Brazilian foundations CNPq for support.

\end{document}